\documentclass[smallextended]{svjour3}       
\smartqed  
\usepackage{graphicx}
\usepackage{amsmath,amssymb,mathrsfs}

\newcommand{\beq}{\begin{equation}}
\newcommand{\eeq}{\end{equation}}
\newcommand{\beqa}{\begin{eqnarray}}
\newcommand{\eeqa}{\end{eqnarray}}
\newcommand{\ba}{\begin{array}}
\newcommand{\ea}{\end{array}}

\begin{document}

\title{Bright solitons in ultracold atoms}

\author{L. Salasnich}

\institute{L. Salasnich \at
Dipartimento di Fisica e Astronomia ``Galileo Galilei'' 
and CNISM, Universit\`a di Padova, via Marzolo 8, 35131 Padova, Italy \\
CNR-INO, via Nello Carrara, 1 - 50019 Sesto Fiorentino, Italy \\
\email{luca.salasnich@unipd.it}}

\date{Received: date / Accepted: date}

\maketitle

\begin{abstract}
We review old and recent experimental and theoretical 
results on bright solitons in Bose-Einstein condensates 
made of alkali-metal atoms and under external optical confinement. 
First we deduce the three-dimensional Gross-Pitaevskii equation (3D GPE) 
from the Dirac-Frenkel action of interacting identical bosons 
within a time-dependent Hartree approximation. Then we discuss 
the dimensional reduction of the GPE from 3D to 1D, deriving 
the 1D GPE and also the 1D nonpolynomial Schr\"odinger equation (1D NPSE). 
Finally, we analyze the bright solition solutions of both 1D GPE and 1D NPSE 
and compare these theoretical predictions with the available experimental data. 

\end{abstract}

\section{Introduction} 

In 1995 three experimental groups achieved Bose-Einstein condensation (BEC), 
i.e. the macroscopic occupation of a single-particle quantum state,
cooling very dilute gases of $^{87}$Rb \cite{cornell}, 
$^7$Li \cite{hulet}, and $^{23}$Na  \cite{ketterle} atoms. 
For these systems the BEC critical temperature is about 
${ T_c}\simeq 100$ nanoKelvin and the gas made of alkali-metal atoms 
is in a meta-stable state which can survive for minutes. 
Another ground-breaking result with ultracold atoms was achieved 
some years later: a stationary optical lattice which traps ultracold atoms 
was obtained with counter-propagating laser beams inside an optical 
cavity \cite{bloch1}. The resulting potential confines 
neutral atoms in the minima of the lattice due to the 
electric dipole of atoms \cite{bloch2}. 
Nowadays the study of neutral atoms trapped with light is a very hot topic 
of research because, changing the intensity and shape of the optical lattice, 
it is possible to confine atoms in very different configurations. 
One can have many atoms per site but also one atom per site \cite{morsch}. 

The main theoretical tool for the study a pure BEC in ultracold and dilute 
alkali-metal atoms is the Gross-Pitaevskii equation \cite{gross}, 
that is a nonlinear Schr\"odinger equation with cubic nonlinearity. 
In 1972 Shabat and Zakharov \cite{sz} found that the 1D cubic 
nonlinear Schr\"odinger equation admits solitonic 
(i.e. shape invariant) analytical solutions. If the 1D nonlinear 
strength is repulsive (self-defocusing nonlinearity) one finds 
the localized dark solitons, while if the nonlinear strength is 
attractive (self-focusing nonlinearity) one finds bright solitons. 
Quite remarkably, both dark and bright solitions have been observed 
experimentally with atomic BECs (see \cite{book-nbec} for a 
comprehensive review). In this paper we concentrate on bright solitons, 
discussing theoretical and experimental results of this exciting 
field of research. 

\section{Gross-Pitaevskii equation} 

Static and dynamical properties of a pure BEC made 
of dilute and ultracold atoms are very well described by the 
Gross-Pitaevskii equation \cite{gross} 
\beq 
i \hbar {\partial\over \partial t} { \psi({\bf r},t)} = 
\left[ -{\hbar^2\over 2m} \nabla^2 + { U({\bf r})} + (N-1) 
{4\pi \hbar^2 { a_s} \over m} |{ \psi({\bf r},t)}|^2  
\right] { \psi({\bf r},t)} \; , 
\label{tdgpe3d}
\eeq
where ${ U({\bf r})}$ is the external trapping potential, $m$ is the mass 
of each atom, and ${ a_s}$ is the s-wave scattering length 
of the inter-atomic potential. 
In this equation ${ \psi({\bf r},t)}$ is the wavefunction of the BEC 
normalized to one, i.e. 
\beq 
\int |{ \psi({\bf r},t)}|^2 \ d^3{\bf r} = 1 \; , 
\eeq
and such that $\rho({\bf r}) = N |{ \psi({\bf r},t)}|^2$ 
is the local number density of the $N$ condensed atoms. 

The Gross-Pitaevskii equation (GPE) can be deduced 
from the many-body quantum Hamiltonian of $N$ identical spinless 
particles 
\beq 
{\hat H} = \sum_{i=1}^N \left( 
-{\hbar^2\over 2m}\nabla_i^2 + { U({\bf r})} \right) + {1\over 2} 
\sum_{\substack{i,j=1\\i\neq j}}^N { V({\bf r}_i - {\bf r}_j)} \; , 
\eeq
where ${ V({\bf r}-{\bf r}')}$ is the inter-atomic potential.  
The time-dependent Schr\"odinger equation of this many-body system is given by 
\beq 
i \hbar  {\partial\over \partial t} {\Psi({\bf r}_1,...,{\bf r}_N,t)} 
= {\hat H} \ {\Psi({\bf r}_1,...,{\bf r}_N,t)} , 
\eeq
where $\Psi({\bf r}_1,...,{\bf r}_N,t)$ is the time-dependent 
many-body wavefuction. 
This time-dependent many-body Schr\"odinger equation is the 
Euler-Lagrange equation of the following many-body Dirac-Frenkel 
\cite{df} action functional 
\beq 
S = \int dt \ d^3{\bf r}_1 \ ... \ d^3{\bf r}_N \ 
\Psi^*({\bf r}_1,...,{\bf r}_N,t)
\left( i\hbar  {\partial\over \partial t} - {\hat H} \right) 
\Psi({\bf r}_1,...,{\bf r}_N,t) \; . 
\eeq

In the case of a pure Bose-Einstein condensate one assumes 
all bosons in the same time-dependent single-particle 
orbital (i.e. a time-dependent version of the 
Hartree approximation \cite{hartree})  
\beq 
{\Psi({\bf r}_1,...,{\bf r}_N,t)} = 
\prod_{i=1}^N { \psi({\bf r}_i,t)} \; . 
\eeq
Inserting this ansatz into the many-body action functional one gets 
\beqa 
S &=& N \int dt \ d^3{\bf r} \ { \psi^*({\bf r},t)} 
\Big( i\hbar  {\partial\over \partial t} + 
{\hbar^2\over 2m} \nabla^2 - { U({\bf r})} 
\nonumber 
\\
&-& {N-1\over 2} \int d^3{\bf r}' \ |{ \psi({\bf r}',t)}|^2 
{ V({\bf r}-{\bf r'})} \Big) { \psi({\bf r},t)}  \; .
\eeqa
The Euler-Lagrange equation of the previous action functional reads 
\beq 
i \hbar {\partial\over \partial t} { \psi({\bf r},t)} = 
\left[ -{\hbar^2\over 2m} \nabla^2 + { U({\bf r})} + (N-1) 
\int d^3{\bf r}' \ |{ \psi({\bf r}',t)}|^2 
{ V({\bf r}-{\bf r'})} \right] { \psi({\bf r},t)} \; .  
\label{eq-hartree}
\eeq
This is the time-dependent Hartree equation for $N$ identical bosons 
in the same single-particle state ${ \psi({\bf r},t)}$. 

In the case of dilute gases one usally assumes (Fermi 
pseudopotential \cite{fermi}) that 
\beq 
V({\bf r}) \simeq g \ \delta^{(3)}({\bf r}) 
\eeq
with $\delta^{(3)}({\bf r})$ the Dirac delta function and, by construction, 
\beq 
g = \int { V({\bf r})} \ d^3{\bf r} \; . 
\eeq
From 3D scattering theory, the s-wave scatering length $a_s$ 
of the inter-atomic potential can be written 
(Born approximation \cite{born}) as 
\beq 
a_s = {m\over 4\pi\hbar^2} \int { V({\bf r})} \ d^3{\bf r} \; . 
\eeq
In this way, from Eq. (\ref{eq-hartree}) one obtains the time-dependent 
3D GPE, Eq. (\ref{tdgpe3d}). 

\section{Dimensional reduction: from 3D to 1D} 

From the Hartree equation (\ref{eq-hartree}) we have obtained 
the time-dependent 3D GPE. 
Clearly, this is the Euler-Lagrange equation of the GP action functional 
\beq 
S = N \int dt \ d^3{\bf r} \ { \psi^*({\bf r},t)} 
\Big( i\hbar  {\partial\over \partial t} + 
{\hbar^2\over 2m} \nabla^2 - { U({\bf r})} 
- {N-1\over 2} { g} |{ \psi({\bf r},t)}|^2 \Big) 
{ \psi({\bf r},t)}  \; .
\label{2-c8}
\eeq

Let us now consider a very strong harmonic confinement of frequency 
${ \omega_{\bot}}$ along $x$ and $y$ and a generic confinement 
${ {\cal U}(z)}$ along $z$, namely 
\beq 
{ U({\bf r})} = 
{1\over 2} m { \omega_{\bot}}^2 (x^2 + y^2) + 
{ {\cal U}(z)} \; . 
\eeq

On the basis of the chosen external confinement, we adopt the ansatz 
\beq 
{ \psi({\bf r},t)} = f(z,t) {1 \over \pi^{1/2} { a_{\bot}}} 
\exp{\left( x^2+y^2 \over 2{ a_{\bot}}^2 \right)} \; , 
\label{3-c8}
\eeq 
where $f(z,t)$ is the axial wave function and 
${ a_{\bot}}=\sqrt{\hbar/(m{ \omega_{\bot}})}$ 
is the characteristic length of the transverse harmonic confinement. 
By inserting Eq. (\ref{3-c8}) into the GP action (\ref{2-c8}) and 
integrating along $x$ and $y$, the resulting effective action functional 
depends only on the field $f(z,t)$.

One easily finds that the Euler-Lagrange equation of the 
axial wavefunction $f(z,t)$ reads 
\beq 
i\hbar {\partial \over \partial t} f(z,t) = 
\left[-{\hbar^2 \over 2m} {\partial^2 \over \partial z^2} + 
{ {\cal U}(z)} + { \gamma} |f(z,t)|^2 \right] f(z,t) \; ,  
\label{1dgpe-c8}
\eeq 
where 
\beq 
{ \gamma}  = {(N-1) { g}\over 2\pi { a_{\bot}}^2} 
\eeq 
is the effective one-dimensional interaction strength and 
the additive constant $\hbar { \omega_{\bot}}$ has been omitted 
because it does not affect the dynamics. 

\subsection{Bright solitons of 1D GPE} 

In the absence of axial confinement, i.e. ${ {\cal U}(z)}=0$, 
the 1D GPE becomes 
\beq 
i\hbar {\partial \over \partial t} f(z,t) = 
\left[-{\hbar^2 \over 2m} {\partial^2 \over \partial z^2} 
+ { \gamma} |f(z,t)|^2 \right] f(z,t) \; .   
\eeq
This is a 1D nonlinear Schr\"odinger equation with cubic nonlinearity. 
In 1972 Shabat and Zakharov \cite{sz} found that 
this equation admits solitonic solutions, such that 
\beq 
f(z,t) = \phi(z-vt) \ e^{i (m v z - m v^2 t/2 - \mu t)/\hbar} \; , 
\eeq 
where $v$ is the arbitrary velocity 
of propagation of the solution, which has a { shape-invariant} axial 
density profile: 
\beq 
\rho(z,t) = N |f(z,t)|^2 = N | \phi(z-vt)|^2 \; . 
\eeq
Setting $\zeta = z-vt$, the 1D stationary GP equation 
\beq 
\left[-{\hbar^2 \over 2m} {d^2 \over d\zeta^2} 
+ { \gamma} |\phi(\zeta)|^2 \right] \phi(\zeta) = 
\mu \ \phi(\zeta) \; ,  
\eeq 
with ${ \gamma} <0$ (self-focusing), 
admits the { bright-soliton solution} 
\beq 
\phi(\zeta) =  \sqrt{ m |{ \gamma}| \over 8\hbar^2} \; 
Sech\left[ {m |{ \gamma}| \over 4\hbar^2}\zeta\right] 
\eeq
with $Sech[x]={2\over e^{x}+e^{-x}}$ and 
\beq 
\mu = - {m \ { \gamma}^2\over 16\ \hbar^2} \; . 
\eeq
In Fig. 1 we plot the  $|\phi(\zeta)|^2$ of the bright soliton 
for three values of the nonlinear strength $\gamma$.

\begin{figure}
\begin{center}
\includegraphics[width=8cm,clip=]{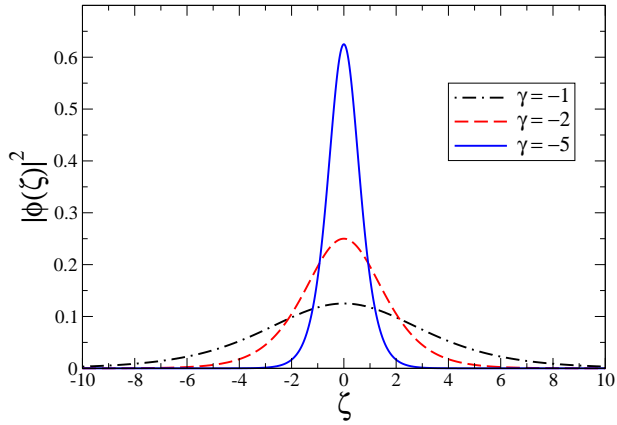}
\end{center}
\caption{Probability density $|\phi(\zeta)|^2$ of the { bright soliton} 
for three values of the nonlinear strength ${ \gamma}$. 
We set $\hbar=m=1$.} 
\end{figure} 

Shabat and Zakharov \cite{sz} used the inverse scattering method \cite{segur} 
to find the explicit expression of $\phi(\zeta)$. Here we use a much simpler 
(but less general) method. 
Let us assume that $\phi(\zeta)$ is real. Then the 1D stationary 
Gross-Pitaevskii equation can be rewritten as 
\beq 
\phi''(\zeta) = - {\partial W(\phi) \over \partial \phi} \; , 
\eeq
where 
\beq 
W(\phi) = {1\over 2} {m |{ \gamma}|\over \hbar^2} \phi^4 + 
{m \mu \over \hbar^2} \phi^2 \; .  
\eeq

Thus, $\phi(\zeta)$ can be seen as the ``coordinate'' for a 
fictitious particle at ``time'' $\zeta$. The constant of motion 
of the problem reads 
\beq 
K = {1\over 2} \phi'(\zeta)^2 + W(\phi) \; , 
\eeq
from which one finds 
\beq 
{d\phi \over d\zeta} = \sqrt{2(K - W(\phi))} \; . 
\eeq

Imposing that $\phi(\zeta)\to 0$ as $|\zeta|\to \infty$ one gets $K=0$ and 
consequently 
\beq 
{d\phi\over \sqrt{- 2 W(\phi)} } = d\zeta \; , 
\eeq
or explicity 
\beq 
{d\phi \over \sqrt{ - {m |{ \gamma}|\over \hbar^2} \phi^4 + 
{2m |\mu| \over \hbar^2} \phi^2} } = d\zeta \; ,  
\eeq
with $\mu <0$. Inserting the integrals one obtains
\beq 
\int_{\phi(0)}^{\phi(\zeta)} 
{d\phi \over \sqrt{ - {m |{ \gamma}|\over \hbar^2} \phi^4 + 
{2m |\mu| \over \hbar^2} \phi^2} } = \zeta  \; . 
\label{bulaba}
\eeq
Setting $\phi'(0)=0$, from the definition of $K$ and using $K=0$ 
one finds $W(\phi(0))=0$ and therefore  
\beq 
\phi(0) = \sqrt{2|\mu|\over |{ \gamma}|} \; . 
\eeq

After integration of Eq. (\ref{bulaba}) one gets 
\beq 
{1\over \sqrt{m |\mu| \over \hbar}} 
ArcSech\left[\sqrt{|{ \gamma}|\over 2|\mu|} \phi(\zeta) \right] = \zeta 
\eeq
from which 
\beq 
\phi(\zeta) = \sqrt{2|\mu|\over |{ \gamma}|} 
Sech\left[\sqrt{m |\mu|\over \hbar^2} \zeta\right] \; . 
\eeq
Finally, imposing the normalization condition 
\beq 
\int dz \ \phi(\zeta)^2 = 1 \; , 
\eeq
one obtains 
\beq 
\mu = - {m \ { \gamma}^2 \over 16 \ \hbar^2}  \; . 
\eeq

\section{Improved dimensional reduction: the 1D NPSE} 

The bright soliton analytical solution has been obtained from the 
1D GPE, which is derived from the 3D GPE assuming a 
transverse Gaussian with a constant transverse width $a_{\bot}$. 
A more general assumption \cite{sala-npse1,sala-npse2,sala-npse3} 
is based on a space-time dependent transverse width 
\beq 
{ \psi({\bf r},t)} = f(z,t) {1 \over \pi^{1/2} 
{ a_{\bot}}{\ \eta(z,t)}} 
\exp{\left( x^2+y^2 \over 2{ a_{\bot}}^2 {\ \eta(z,t)}^2 
\right)} \; , 
\eeq 
where $f(z,t)$ is the axial wave function and 
${\ \eta(z,t)}$ is the {\ adimensional transverse width} 
in units of ${ a_{\bot}}$. 
From this ansatz one gets the {1D nonpolynomial Schr\"odinger 
equation} (1D NPSE) 
\beqa
i\hbar {\partial \over \partial t} f &=& 
\left[-{\hbar^2 \over 2m} {\partial^2 \over \partial z^2} + 
{ {\cal U}(z)} 
+ {{ \gamma} |f|^2 \over {\ \eta}^2} + {\hbar
{ \omega_{\bot}} \over 2} 
\left( {1\over {\ \eta}^2} + {\ \eta}^2 \right) 
\right] f \; ,    
\\
{\ \eta} &=& \left( 1 + { \gamma} |f|^2\right)^{1/4} \; . 
\eeqa
In the weak-coupling regime ${ \gamma} |f|^2\ll 1$ one finds 
${\ \eta} \simeq 1$ and the 1D NPSE becomes the familiar 1D GPE. 

\subsection{Bright solitons of 1D NPSE}

With ${ {\cal U}(z)}=0$ and assuming ${ \gamma} <0$ 
the NPSE admits analytical { bright soliton} solutions. Setting 
\beq 
f(z,t) = \phi(z-vt) e^{i (mv^2/2 - \mu )t/\hbar} , 
\eeq 
one finds the { bright-soliton} solution written in implicit form 
\beqa 
\zeta= {1\over \sqrt{2}} {1\over \sqrt{1-\mu}} \; 
arctg\left[ 
\sqrt{ \sqrt{1-|{ \gamma}|\phi^2}-\mu \over 1-\mu } 
\right] 
\\
-{1\over \sqrt{2}} {1\over \sqrt{1+\mu}} \; 
arcth\left[ 
\sqrt{ \sqrt{1-|{ \gamma}|\phi^2}-\mu \over 1+\mu } 
\right] \; ,  
\eeqa 
where $\zeta =z-vt$ and $|{ \gamma}|=2|{ a_s}|(N-1)/
{ a_{\bot}}$. 

\begin{figure}
\begin{center}
\includegraphics[width=9cm,clip=]{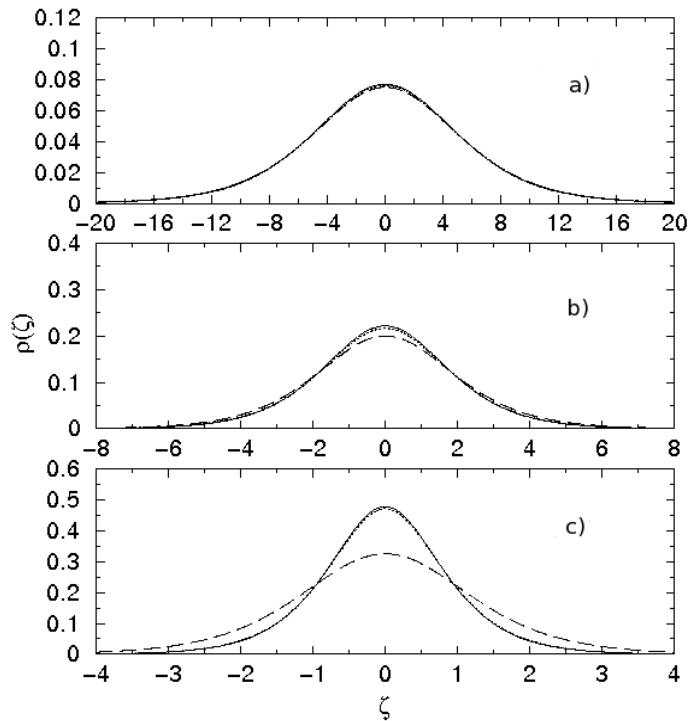}
\end{center}
\caption{Axial probability density $\rho(\zeta)=|\phi(\zeta)|^2$ 
of the Bose-condensed bright soliton: 
3D GPE (full line), 1D NPSE (dotted line), 1D GPE (dashed line). 
Length in units ${a_{\bot}}=(\hbar /m{\omega_{\bot}})^{1/2}$ 
and density in units $1/{a_{\bot}}$. 
Three values of the interaction strength: 
a) ${\gamma}=0.3$, b) ${\gamma}=0.8$, c) ${\gamma}=1.3$. 
Adaped from \cite{sala-npse2}. } 
\end{figure} 

Fig. 2 reports the axial probability density $\rho(\zeta)=|\phi(\zeta)|^2$ 
of the bright soliton obtained by using the 3D GPE (full line), 
the 1D NPSE (dotted line), and the 1D GPE (dashed line). 
In the weak-coupling limit (${ \gamma}\phi^2 \ll 1$) one finds that 
the NPSE bright-soliton solution reduces the the 1D GPE one. 
This is clearly shown in the figure. 
However, contrary to the 1D GPE bright soliton, 
the 1D NPSE bright soliton does not exist 
anymore, {collapsing} to a Dirac delta function, at 
$$
{ \gamma_c} 
= \left( {2{ a_s}(N-1)\over { a_{\bot}}} \right)_c 
= - {4\over 3} \; . 
$$
This analytical result is in extremely good agreement with the 
numerical solution of the 3D GPE \cite{sala-npse2}. Indeed, 
Fig. 2 shows that up to the collapse the density profile 
obtained with the NPSE is very close to the 3D GPE one. 

\section{Bright solitons in experiments with ultracold atoms} 

In 2002 there were two relevant experiments \cite{exp-sol1,exp-sol2} 
about bright solitons with BECs made of $^7$Li atoms. 
Both experiments used the technique of Fano-Feshbach resonance \cite{ff}
to tune the s-wave scattering length $a_s$ of the inter-atomic 
potential by means of an external constant magnetic field. 

\begin{figure}
\begin{center}
\includegraphics[width=8cm]{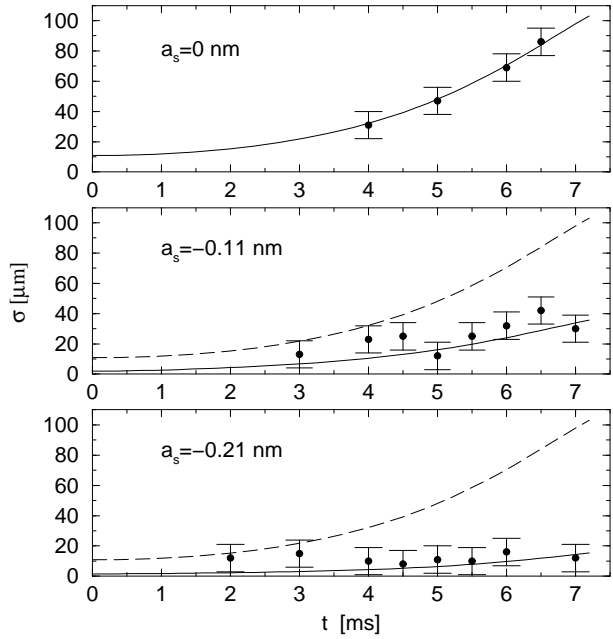}
\end{center}
\caption{Root mean square size $\sigma$ of the longitudinal width 
of the BEC vs propagation time $t$. $a_s$ is the s-wave scattering length 
of the inter-atomic potential, which is experimentally tuned 
by means of a Fano-Feshbach resonance \cite{ff}. Filled circles are 
experimental data taken from \cite{exp-sol1}. The dashed 
line is the ideal gas ($a_s=0$) curve. 
The solid line is obtained solving the time-dependent 
NPSE. Adapted from \cite{sala-ens}.} 
\end{figure}

Khaykovich et al. \cite{exp-sol1} reported the production of 
bright solitons in an ultracold $^7$Li gas. The interaction 
was tuned with a Feshbach resonance from repulsive to attractive 
before release in a one-dimensional optical waveguide, which is attractive 
in the transverse direction but expulsive in the longitudinal direction.  
Propagation of the soliton without dispersion over a macroscopic 
distance of $1.1$ millimeter was observed in the case of attractive 
interaction. In their experiment, Khaykovich {\it et al.} \cite{exp-sol1} 
measured the root mean square size $\sigma$ of the longitudinal 
width versus the propagation time for three values of $a_s$: 
$a_s=0$, $a_s=-0.11$ nm, and $a_s=-0.21$ nm. 
Fig. 3 shows the experimental 
data \cite{exp-sol1} and our numerical 
results obtained with the time-dependent 1D NPSE \cite{sala-ens}. 
The agreement between experiment and theory is quite good. 
The numerical results are obtained by using a finite-difference 
Crank-Nicolson scheme with predictor-corrector \cite{sala-numerics}. 
Very recently, various open access and high performance 
codes have been developed for the solution of the 
time-dependent GPE \cite{antun-numerics}.

Strecker et al. \cite{exp-sol2} reported the formation of a train of 
bright solitons of $^7$Li  atoms in a quasi-one-dimensional optical trap 
by a sudden change in the sign of the scattering length 
from positive to negative. The solitons were set in motion 
by offsetting the optical potential, 
and were observed to propagate in the longitudinal harmonic 
potential for many oscillatory cycles without spreading. 

\begin{figure}
\begin{center}
\includegraphics[width=8cm]{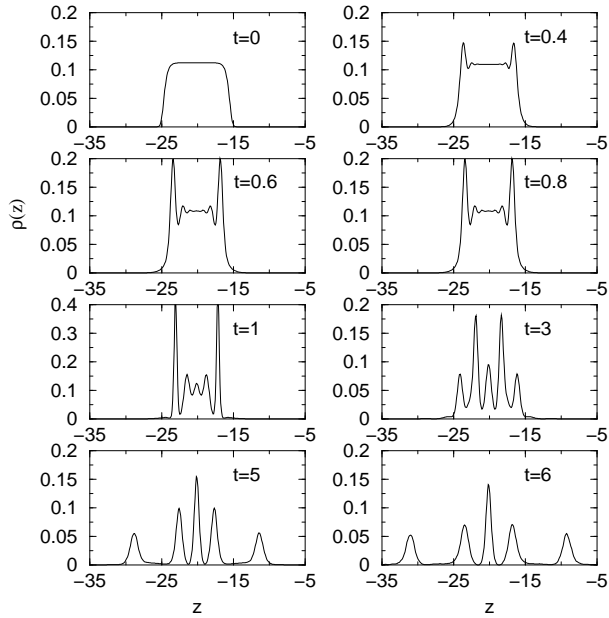}
\end{center}
\caption{Axial density profile $\rho(z)$ of the 
BEC made of $10^4$ $^7$Li atoms obtained by solving the 3D GPE. 
For $t<0$ the scattering length is $a_s=100a_B$, while 
for $t\geq 0$ we set $a_s=-3a_B$ with $a_B$ the Bohr radius. 
Length $z$ is in units of the characteristic 
length $a_z=\sqrt{\hbar/(m\omega_z)}$
of the weak axial harmonic confinement of frequency $\omega_z$. 
Time $t$ in units of $1/\omega_z$. Density $\rho$ in units of $1/a_z$. 
Adapted from \cite{sala-rice}.} 
\end{figure}

We successfully simulate the soliton train formation of Ref. \cite{exp-sol2} 
by using the time-dependent 3D GPE \cite{sala-rice}. 
In Fig. 4 we plot the probability density 
in the longitudinal direction $\rho(z)$ of the BEC 
made of $10^4$ $^7$Li atoms. Initially there is 
a stable condensate of $^7$Li atoms with a 
large positive scattering length $a_s$ but at time $t=0$ 
the scattering length $a_S$ is switched to a negative value. 
A number $N_S$ of bright solitons is produced and this can be interpreted 
in terms of the modulational instability of the 
time-dependent macroscopic wave function of the Bose condensate \cite{lc}
An estimate of the number $N_s$ of bright solitons which 
are generated is 
\beq 
N_s = {\sqrt{ N |a_s| L} \over \pi a_{\bot} } \; , 
\eeq
where $a_s$ is the final negative scattering length, $N$ is the total 
number of atoms, $a_{\bot}$ is the characteristic length of transverse 
harmonic confinement and $L$ is the initial longitudinal length 
of the quasi-1D BEC \cite{sala-rice}. This formula, 
based on the analysis of the imaginary Bogoliubov spectrum 
of elementary excitations (see \cite{sala-rice} for details), 
is in good agreement with both experimental results and numerical simulations. 

Very recently, the formation of matter-wave soliton trains 
by modulational instability was experimentally reexamined by 
Nguyen {\it et al.} \cite{exp-sol3}. They used a nearly nondestructive 
imaging technique to follow the dynamics of thess trains finding that 
the modulation instability is driven by noise and neighboring solitons 
interact repulsively during the initial formation of the soliton train. 
These findings are indeed in full agreement with our theoretical 
predictions based on the numerical simulation of 3D GPE 
and 1D NPSE \cite{sala-rice}. 

\section{Conclusions} 

In this paper we have explicitly derived the analytical solution of 
the 1D bright soliton from the one-dimensional Gross-Pitaevskii 
equation (1D GPE), which, in turn, is obtained 
from the 3D GPE assuming a transverse Gaussian with a constant width
${ a_{\bot}}$. We have then shown that a more general assumption, with 
a space-time dependent transverse width, gives rise the 1D 
nonpolynomial Schr\"odinger equation (1D NPSE). 
1D NPSE admits bright solitons which collapse at a critical 
interaction strength, in good agreement with the findings of full 3D GPE. 
Both 3D GPE and 1D NPSE are reliable tools 
to reproduce the available experimental data \cite{exp-sol1,exp-sol2,exp-sol3} 
of BEC bright solitons made of alkali-metal atoms. The experimental 
study of bright solitons in ultracold atoms is still a hot topic, 
as is evident considering the very recent experiments on the train of 
bright solitons \cite{exp-sol3} and on attractive two-component 
bosonic mixtures \cite{exp-sol4}. Moreover, in the last few years, 
it is has been suggested that atomic bright solitons 
can be produced with attractive two-component 
mixtures \cite{sala-mix,petrov}, with space-dependent scattering 
lengths \cite{boris-nlo}, and also with artificial spin-orbit and Rabi 
couplings \cite{altri-spin,sala-spin,altri2-spin}. 
These remarkable theoretical predictions 
need experimental confirmation. 


\end{document}